\documentclass{emulateapj}
\usepackage{graphicx} 
\usepackage{epstopdf} 
\usepackage{amsmath}
\usepackage{hyperref}
\usepackage{natbib}

\newcommand{\sub}[2]{\ensuremath{#1_{\mathrm{#2}}}}
\newcommand{\super}[2]{\ensuremath{#1^{\mathrm{#2}}}}
\newcommand{\unit}[2]{\ensuremath{\textrm{#1}^{#2}}}
\newcommand{\half}{\frac{1}{2}}
\newcommand{\vect}[1]{\mathbf{#1}}

\shorttitle{Modeling with streams in cosmological halos}
\shortauthors{Sanderson, Hartke, \& Helmi 2015}

\begin{document}
\title{Modeling the gravitational potential of a cosmological dark matter halo\\ with stellar streams}

\author{Robyn E. Sanderson\altaffilmark{1}}
\altaffiltext{1}{NSF Astronomy and Astrophysics Postdoctoral Fellow}
\email{robyn@astro.columbia.edu}
\affil{Department of Astronomy, Columbia University, 550 W 120th St, New York, NY 10027, USA}
\affil{Kapteyn Astronomical Institute, University of Groningen, P.O. Box 800, 9700 AV Groningen, Netherlands}

\and
\author{Johanna Hartke \& Amina Helmi}
\affil{Kapteyn Astronomical Institute, University of Groningen, P.O. Box 800, 9700 AV Groningen, Netherlands}
 
 \slugcomment{version revised \today}

\begin{abstract}
Stellar streams result from the tidal disruption of satellites and star clusters as they orbit a host galaxy, and can be very sensitive probes of the gravitational potential of the host system. We select and study narrow stellar streams formed in a Milky-Way-like dark matter halo of the Aquarius suite of cosmological simulations, to determine if these streams can be used to constrain the present day characteristic parameters of the halo's gravitational potential. We find that orbits integrated in static spherical and triaxial NFW potentials both reproduce the locations and kinematics of the various streams reasonably well. To quantify this further, we determine the best-fit potential parameters by maximizing the amount of clustering of the stream stars in the space of their actions. We show that using our set of Aquarius streams, we recover a mass profile that is consistent with the spherically-averaged dark matter profile of the host halo, although we ignored both triaxiality and time evolution in the fit. This gives us confidence that such methods can be applied to the many streams that will be discovered by the Gaia mission to determine the gravitational potential of our Galaxy.
\end{abstract}

\section{Introduction}
\label{sec:intro}
Stellar streams are the result of the accretion of globular clusters or dwarf galaxies onto a more massive host galaxy, and are formed as their stars are tidally stripped under the influence of the potential of the host. The knowledge of a stream's trajectory provides a constraint on the potential and thus on the matter distribution and shape of the dark halo \citep{2009BinneyA, 2010LawA, 2009Willett, 2010Newberg, 2013MNRAS.433.1826S}. 

There exist many streams in the Milky Way halo; among the most
well-studied is the stream associated with the Sagittarius dwarf
galaxy \citep{1994Ibata,2000Ivezic, 2000Yanny}. This stream is a good
example of how assumptions about the symmetry, shape, and functional
form of the Milky Way's dark halo, combined with incomplete knowledge
of the phase-space of stars in the tidal stream, can lead to
conflicting conclusions about the mass distribution of the Milky Way
(MW). Based on 3D positions and radial velocities (a total of 4
phase-space coordinates) of about 75 carbon stars, \citet{2001Ibata}
argued that the dark halo should be nearly spherical since the stars'
positions roughly followed a great circle on the sky.  Analysis of
parts of the stream discovered with the Sloan Digital Sky Survey
suggested that the mass distribution could be oblate
\citep{2004Martinez}, and a similar conclusion was reached using the
precession of the orbital plane of the stream's M giants from 2MASS
\citep{2005Johnston}. On the other hand, the radial velocities of the
leading stream's M giants from 2MASS clearly favored a
prolate shape \citep{2004Helmi}. These works used the 4D phase-space
coordinates of several hundred stars. The conundrum was solved by
\citet{2010LawA} who showed that a triaxial halo could reconcile
the angular position of the stream with its radial velocities. In
their model, a logarithmic density profile was assumed with axis
ratios and orientation constant with radius, leading to a best fit
where the disk was parallel to the intermediate axis of the dark halo;
a dynamically untenable situation. More recently,
\citet{2013Vera-Ciro} argued that it is not necessary to assume a constant shape with radius, and that
there is enough freedom in the data to allow for a model that is oblate
and aligned with the disk at small radii. Furthermore, the
authors find that if the gravitational contribution of the Large
Magellanic Cloud is included, the resulting best-fit Milky Way halo 
resembles those in cosmological simulations at large radii. The \citet{2010LawA} model in
the outskirts may be seen as an effective potential, which is the sum
of that of the LMC and of the underlying halo of the Galaxy. Although 
\citet{2013Ibata} argue that a spherical halo with an unusual rising rotation curve out
to nearly 50 kpc can fit the data, the velocities are clearly less well reproduced in this model.  
\citet{Belokurov2014} also argue that the azimuthal precession rate can be used to measure the radial dependence of the 
mass distribution. Clearly, thus far not all of the available data has been used optimally nor has the modelling been sufficiently general 
from assumptions about the mass distribution: its
shape, radial profile, and even the effect of substructures like the
LMC, to reach its full potential.

How then can the mass distribution of the MW's dark halo be unambiguously determined? One could for example try to fit more than one stream, or try to obtain the remaining phase-space coordinates for the stream being fitted, in the hopes that this could break the degeneracy. Given that stars in a single stream all have similar orbits, one would expect that even perfect and complete data on a single stream would be insufficient to break all the degeneracies. It is unclear how many streams, or which ones, would be sufficient to do this. However, the Gaia mission \citep{2001Perryman}, launched in December 2013, will at least make it possible to analyze multiple streams simultaneously, by measuring the positions and velocities of 1 billion Milky Way stars, including many halo stars. Gaia will measure full six-dimensional phase-space coordinates for roughly 15 percent of these stars, and five-dimensional coordinates for the remaining 85 percent. This dataset will likely contain hundreds of streams \citep{1999Helmi} and its uniformity will enable simultaneous analysis of multiple streams. 

Even in this case, however, the fact would remain that the MW's dark halo does not precisely follow a particular functional form; it is clear from the example of the Sgr stream that our assumptions do affect the results of the fit. In this work we explore how Gaia's upcoming observations can help resolve some of these problems, and test how simplifying assumptions about the potential are projected onto both the ability of orbits to resemble a given stream, and the resulting best-fit profile. 

Since many stream-fitting algorithms \citep[e.g.,][]{2011MNRAS.417..198V,2014MNRAS.443..423S,2014ApJ...794....4P,2014ApJ...795...94B} compare the positions and velocities of stream stars to models in 6-dimensional phase space (or less), our first goal is to determine whether a center-of-mass orbit integrated in an spherical instead of triaxial potential lined up equally well (or badly) for streams on different kinds of orbits. Though streams do not exactly follow orbits \citep[e.g.][]{2009BinneyA}, the degree to which an integrated orbit lines up with the stars of a stream is a simple proxy for whether a given potential will be able to produce a stream that fits the observations; in fact to save computing time the best-fit single orbit is often used as a starting point for the progenitor's center-of-mass orbit when searching parameter space with N-body simulations  \citep[e.g.][]{fardal:2006aa,2010Law}. We wish to determine whether this strategy will be effective for the streams produced self-consistently from satellites in the Aquarius simulations. Additionally, the degree to which streams lie in or out of a plane has been used to conjecture about the spherical symmetry (or lack thereof) of the MW halo \citep[e.g.][]{2001Ibata}. We wish to determine whether the Aquarius streams can be successfully used in this way, and if so, which streams are most sensitive.

Our second goal is to test the robustness of the results of a new potential-fitting algorithm, based on maximizing the information content of the action space of stream stars \citep{2015ApJ...801...98S}, if the potential being fit was substantially less complicated than the real potential. Unlike many methods this one does not directly compare positions and velocities of stream stars with a model, but does analyze multiple streams simultaneously. We want to test whether the results from fitting a simple spherical potential will still reflect the true mass distribution of the halo to the extent permitted by such an oversimplified model.

To conduct the two tests we selected stellar streams produced from the cosmological, dark-matter-only N-body simulation Aquarius A \citep[][hereafter S08]{2008Springel} via stellar tagging according to a semianalytic model of star formation \citep{2010Cooper} and used them to evaluate different potential models. In Section \ref{sec:data}, we describe the two models for the dark halo: a triaxial and a spherical NFW potential both fit to the known dark-matter distribution of the simulated halo. In Section \ref{sec:integration} we selected 15 structures based on their streamy appearance and low mass (i.e. narrow width) and integrated center-of-mass orbits for each stream in the two different potential models, to see how well the orbits traced the streams. Then in Section \ref{sec:KLD} we fit a spherical NFW model simultaneously to all 15 streams using the \citeauthor{2015ApJ...801...98S} action-clustering method and compared the best-fit result  to the spherically averaged DM distribution from the N-body simulation. In Section \ref{sec:discussion} we discuss our results and implications for future work.

\section{Data and models}
\label{sec:data}
\begin{figure*}
   \centering
   \includegraphics[width=.95\hsize]{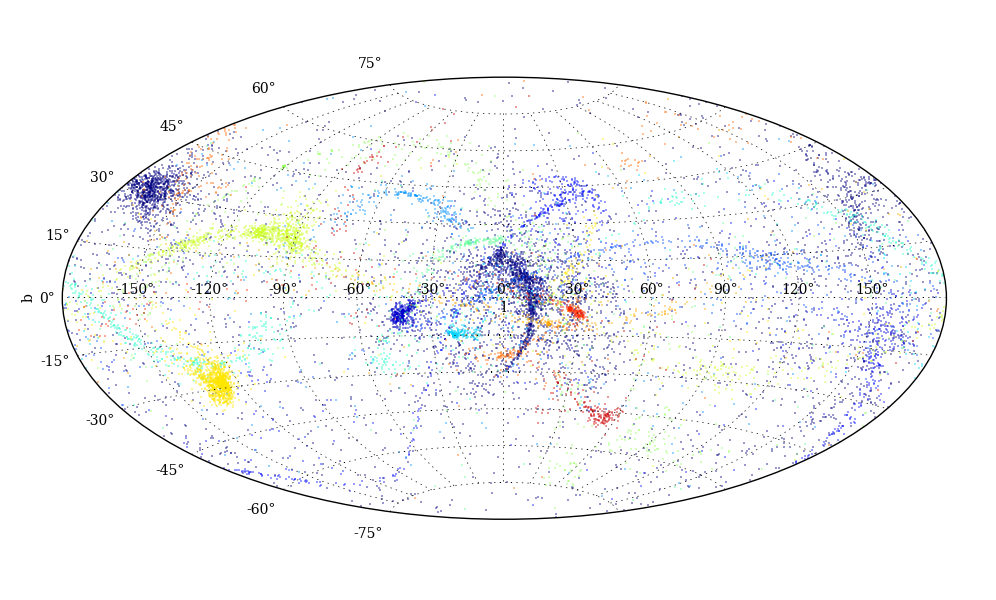}
      \caption{Selected streams projected onto the sky. The Sun is located at 8 kpc along the major (x-) axis of the simulation (see Section \ref{ssec:potentials} for a discussion of the coordinate system). Each stream is represented by a different color.}
         \label{fig:all_streams}
\end{figure*}

The Aquarius project is a suite of N-body simulations of Milky-way sized halos run in a $\Lambda$CDM cosmology (S08). Six different halos, labeled A to F, were simulated, each of them at different resolutions, labeled by the numbers 1 (highest) to 5 (lowest). Stellar populations were associated to subsets of CDM particles \citep{2010Cooper} using the semi-analytic \textsc{Galform} model \citep{2000Cole}. In this work we focus on the halo Aquarius-A-2 (Aq-A-2) at redshift $z = 0$. The mass, shape and orientation of the halo change over time \cite{2011Vera}, and it contains a population of subhalos resolved down to about $10^5$ solar masses (S08). 

From the set of dark matter particles tagged by \citet{2010Cooper}, we select those associated with infalling luminous satellites with total stellar mass between $1\times 10^{3}\;M_{\odot}$ and $5\times 10^{5}\;M_{\odot}$ and which gave rise to structures that appeared spatially coherent or ``streamy"  (i.e. long and thin) in position space \citep{2011Helmi}. The selected streams are shown in Figure \ref{fig:all_streams} as a sky projection viewed from the centre-of-mass of the host halo. Different streams are shown in different colors in this and subsequent plots.

\subsection{Potentials and parameters}
\label{ssec:potentials}
We use the Navarro-Frenk-White (NFW) profile \citep{1996NFW} to describe the mass profile of a dark halo with scale radius $r_s$ and scale density $\rho_s$. In order to avoid degeneracies in action space when using the KLD method (cf. section \ref{sec:KLD}), we use the enclosed mass at the scale radius,
\begin{equation}
M_s \equiv M_{encl}(r_s) = 4\pi\rho_s r_s^3 \left( \ln 2 - \frac{1}{2} \right),
\label{eq:MRsrhos}
\end{equation}
as one of the two parameters in the potential rather than $\rho_s$ (the other parameter is the scale radius $r_s$). In terms of $M_s$ and $r_s$, the potential is
\begin{equation}
	\Phi(r) = -G \frac{M_s}{\ln 2 - 1/2} \frac{\ln (1 + r/r_s)}{r}
\label{eq:PhiMsRs}
\end{equation}
where $G$ is Newton's constant.

For a spherical NFW halo, the radius $r$ is simply defined as $r^2 = x^2 + y^2 + z^2$. To produce a triaxial NFW halo with the same overall mass and scale radius, we follow \citet{2008Vogelsberger} and define the ellipsoidal radius 
\begin{equation}
\label{eq:rEllipsoidal}
r_E^2 \equiv \left(\frac{x}{a}\right)^2 + \left(\frac{y}{b}\right)^2 +  \left(\frac{z}{c}\right)^2,
\end{equation}
where $a$, $b$, and $c$ are the relative lengths of the major, intermediate, and minor axes respectively. In this orientation the major axis of the ellipsoid is therefore aligned with the $x$ axis, and so forth. In order to maintain the proper normalization we require $a^2 + b^2 + c^2 = 3$. We then replace $r$ in the spherical NFW potential with the quantity
\begin{equation}
\label{eq:rTilde}
\tilde{r} = \frac{(r_a + r)r_E}{r_a + r_E},
\end{equation}
where $r_a$ is the scale over which the potential shape transitions from ellipsoidal to near spherical. As in \citet{2008Vogelsberger} we set $r_a=2r_s$. This produces a halo that is ellipsoidal in the center and becomes spherical for $r \gg r_a$.

We use the axis ratios for the potential of Aq-A determined by \cite{2011Vera} using the method for defining isopotential contours described in \cite{2007Hayashi}. The shape of Aq-A changes as a function of radius; we take the axis ratios of the potential at the scale radius from Figure A2 of \cite{2011Vera} ($b/a=0.90$, $c/a=0.85$) and scale the axis lengths to the proper normalization for the ellipsoidal radius, obtaining $a = 1.09$, $b = 0.98$, and $c = 0.93$. The rotation matrix that transforms from the coordinate system of the Aquarius simulations $(x_A,y_A,z_A)$ to the one aligned with the ellipsoidal axes $(x,y,z)$ was determined by \citet{2011Vera}:
\begin{equation}
\begin{pmatrix}
x\\
y\\
z\\
\end{pmatrix}
=
\begin{pmatrix}
0.24 & -0.73 & 0.64\\
-0.12 & 0.63 & 0.76\\
-0.96 & -0.27 & 0.07\\
\end{pmatrix}
\begin{pmatrix}
x_A\\
y_A\\
z_A\\
\end{pmatrix}
\end{equation}
We use this matrix to rotate the coordinates of the selected stream stars.

\begin{deluxetable*}{rlcccc|c}
\tablecaption{Properties of the Aq-A-2 halo.} 
\tablehead{
\colhead{Property} & \colhead{Unit} & \colhead{S08 meas} & \colhead{S08 calc} & \colhead{N10 meas} & \colhead{N10 calc} & \colhead{Best-fit}
}
\startdata
$M_{200}$ & $10^{12} M_\odot$ & 1.842 & -- & 1.842 & $1.322$ & $1.530^{+0.818}_{-0.402}$  \\

$r_{200}$ & kpc & 245.88 & -- & 245.88 & -- & $230.28^{+35.35}_{-22.26}$ \\

$\delta_V$ & $10^4$ & 2.060 & 2.130 & 2.038 & 1.529 & $1.230^{+1.364}_{-0.692}$ \\

$c_{NFW}$ & -- & -- & 16.19 & -- & -- & $13.18^{+4.54}_{-3.76}$ \\

$r_{-2}$ & kpc/$h$ & -- & -- & 11.15 & -- & $12.82^{+3.38}_{-1.82}$ \\

$\rho_{-2}$ & $10^6 h^2 M_\odot$ \unit{kpc}{-3} & -- & 10.199 & 7.332 & -- &  $6.156^{+6.828}_{-3.465}$\\

$\rho_{s}$ & $10^6 M_\odot$ \unit{kpc}{-3} & -- & $21.98$\tablenotemark{1} & -- & $15.78$\tablenotemark{2} &  $13.27^{+14.72}_{-7.47}$\\

$r_s$ & kpc & -- & 15.19 & -- & 15.19 & $17.47\substack{+4.60\\-2.48}$ \\

$M_s$ & $10^{12} M_\odot$ & -- & $0.187$ & -- & $0.134$ & $0.172\substack{+0.057\\-0.020}$ \\

$\sub{r}{max}$ & kpc & 28.14 & 32.87 & 28.14 & 32.87 & $37.77^{+9.95}_{-5.36}$  \\
$\sub{v}{max}$ & km \unit{s}{-1} & 208.49 & 248.97 & 208.49 & 210.96 & $217.44^{+53.58}_{-35.80}$ \\
\enddata
\tablecomments{Properties of Aq-A-2 from different sources: measured directly from the Aq-A simulated halo as described in the text, with no assumption of an NFW profile (columns ``S08 meas" and ``N10 meas"), calculated from the directly-measured quantities assuming the profile is NFW (``S08 calc" and ``N10 calc"), and quantities derived from our best-fit $M_s$ and $r_s$ values. The scaled Hubble constant is $h = 0.734$. \tablenotetext{1}{Calculated using Equation \eqref{eqn:rho1}, with $M_{200}$ and $r_{200}$ as measured and $c$ related to $\delta_V$ via Equation \eqref{eq:deltaV}.} \tablenotetext{2}{Calculated using $\rho_s=4\rho_{-2}$.}}
\label{tab:par}
\end{deluxetable*}

For the mass and scale radius of the Aq-A-2 halo, we have several different options. S08 and \citet[][hereafter N10]{2010Navarro} both determine slightly different sets of halo parameters for Aq-A-2 that lead to different values for $M_s$ assuming a spherical NFW halo; both papers find the same value for $r_s$ under this assumption. 

To obtain a value for $r_s$, S08 determine the radius $r_{200}$ in which the virial mass $M_{200}$ is enclosed (the mass enclosed in a sphere with average density 200 times the critical value). Then, from the peak value of the circular velocity curve $v_{max}$ at $r_{max}$, they determine the characteristic density contrast $\delta_V$, 
\begin{equation}
\delta_V \equiv 2 \left(\frac{\sub{V}{max}}{H_0 \sub{r}{max}}\right)^2,
\end{equation}
which can be converted to the standard NFW concentration $c$ via \citep{1996NFW}
\begin{equation}
\label{eq:deltaV}
	\delta_c = 7.213 \delta_V = \frac{200}{3} \frac{c^3}{\ln(1 + c) - \frac{c}{1 + c}}.
\end{equation}
This results in in $c = 16.19$ for Aq-A-2 assuming that the halo is well fit by an NFW profile. The virial radius is then related to the NFW scale radius by
\begin{equation}
r_s = r_{200}/c = 15.19\;\text{kpc}.
\label{eq:Rsr200c}
\end{equation}
$M_{200}$ and $r_{200}$ are related to the scale density $\rho_s$ via
\begin{eqnarray}
	M_{200} &=& M_{\text{encl}}(r_{200}) = 4\pi\rho_s (r_{200}/c)^3 \left( \ln (1 + c) - \frac{c}{1 + c} \right) \nonumber \\
	\rho_s &=& \frac{M_{200}}{4\pi (r_{200}/c)^3 \left( \ln (1 + c) - \frac{c}{1 + c} \right)}, 
	\label{eqn:rho1}
\end{eqnarray}
or, equivalently, to $\delta_V$ via:
\begin{equation}
	\rho_s = \delta_c \rho_{crit} = 7.213 \delta_V \frac{3 H^2}{8\pi G}.
	\label{eqn:rho2}
\end{equation}
Both relations result in a value for $M_s$ (using Equation \ref{eq:MRsrhos}) of $1.87\times10^{11}\ M_{\odot}$.

On the other hand, N10 characterize the same halo by determining the radius $r_{-2}$ where the logarithmic slope of the profile, $\gamma (r) = -\text{d} \ln \rho / \text{d} \ln r$, equals the isothermal value, $\gamma = 2$. The density at $r_{-2}$ is denoted as $\rho_{-2}$. For a NFW profile, $r_{-2} = r_s$ and $4\rho_{-2} = \rho_s$. The value obtained for $r_s$ this way is identical to that obtained by finding $r_{200}$ and applying Equation \eqref{eq:Rsr200c}, but the value obtained for $M_s$ using this method is somewhat smaller, $1.34\times10^{11} \ M_{\odot}$.

The different results are summarized in Table \ref{tab:par}. The parameters from S08 give the correct $M_{200}$ by definition, but the NFW profile with this $M_s$ and $r_s$ has a significantly higher peak circular velocity than is measured directly from the numerical simulation of the halo. On the other hand, the parameters from N10 give close to the the correct peak of the circular velocity curve, but significantly underestimate the enclosed mass at $r_{200}$. This discrepancy occurs because the Aq-A halo mass profile is not strictly NFW and these two methods normalize the mass profile at two different radii. The N10 profile agrees with the spherical mass profile best within 40 kpc, while the S08 profile is too high until near the virial radius. The streams we will use to fit the halo orbit in a radial range from inside $r_{-2}$ out to about half of $r_{200}$. Thus it is likely that our fit will match the empirical mass profile best over some radial range intermediate to these two, and in Section \ref{sec:KLD} we will compare our fit results to both values of $M_s$   since they bracket the possible masses obtained by fitting an NFW profile to the halo, depending on which range of radii is used for the fit. For integrating orbits, we used the values determined by S08, but the range of masses we explore includes the N10 profile.

Since Aq-A is not well fit by a single NFW profile at all radii, we also determined its radial mass profile directly from the dark matter particle data to compare to our fit results. The empirical mass profile was obtained by first removing all bound substructures identified by the structure finder \textsc{Subfind} \citep[][also used to produce the stellar stream catalog]{2001MNRAS.328..726S}, then binning the remaining particles in spherical radius. Because the bound substructures are removed, this empirical profile has a slightly lower virial mass than S08.

\section{Orbit integration}
\label{sec:integration}

\subsection{Methods}
\label{ssec:integrationMethods}
We integrated center-of-mass orbits for each selected stream from Aq-A-2 in the spherical and triaxial NFW potentials described above, using as initial conditions the position and velocity of a particle chosen by eye to lie about midway along each stream In the case of stream 1051588, we tried using a range of different particles at different positions along the stream, but none of the orbits we integrated traced the stream closely at all. 

The equations of motion were integrated numerically with \texttt{scipy} using  a fourth-order Runge-Kutta algorithm\citep{1988dopri}, with a timestep of 0.01 Gyr for both the spherical and the triaxial potentials. Each orbit was integrated forward and backward in time, starting from the initial conditions, for 2 Gyr in each direction. The center-of-mass orbits in each potential were compared to the current positions of the stream stars, to determine whether the spherical or triaxial potential produced an orbit that more closely followed the stream. Inspired by comparisons used in fitting orbits to streams, we calculated for each stream the minimum distance between the integrated orbit of the central particle and each star in the stream. The phase-space location along the orbit $(\sub{\vect{x}}{orb},\sub{\vect{v}}{orb})$ was tabulated at each timestep using the orbit integration and compared to the phase-space location of each star particle by summing the squares of the minimum distances, in the spirit of a chi-squared. We calculate this statistic independently for position and velocity and normalize by the number of stars in each stream:
\begin{eqnarray}
\chi^2_x &\equiv & \frac{1}{N_*} \sum_{i=0}^{N_*} \textrm{min}\left[ (\sub{\vect{x}}{orb} - \vect{x}_i)^2 \right] \nonumber \\
\chi^2_v &\equiv & \frac{1}{N_*} \sum_{i=0}^{N_*} \textrm{min}\left[ (\vect{v}_{\mathrm{orb}} - \vect{v}_{i})^2\right] \label{eq:chisq}
\end{eqnarray}
To eliminate a few outliers, we discard any stars in the streams whose minimum distance from the orbit is larger than 25 kpc. For most streams no stars are thrown out; for a few of the largest a handful of stars are discarded. The largest number of discarded stars is 66, from stream 1025754 which contains 5158 stars in total. Most of these are well outside this distance, and our results do not change appreciably with the cutoff distance. 

We also explored how changing the parameters $M_s$ and $r_s$ affected the agreement between the integrated orbit and the stream. Having determined which potential (spherical or triaxial) produced the best stream-orbit agreement with the known parameters, we then varied each parameter from 0.25 to 2 times its known value while holding the other fixed. 

\subsection{Results}
\label{ssec:orbits}

\begin{figure}
\includegraphics[width=0.45\textwidth]{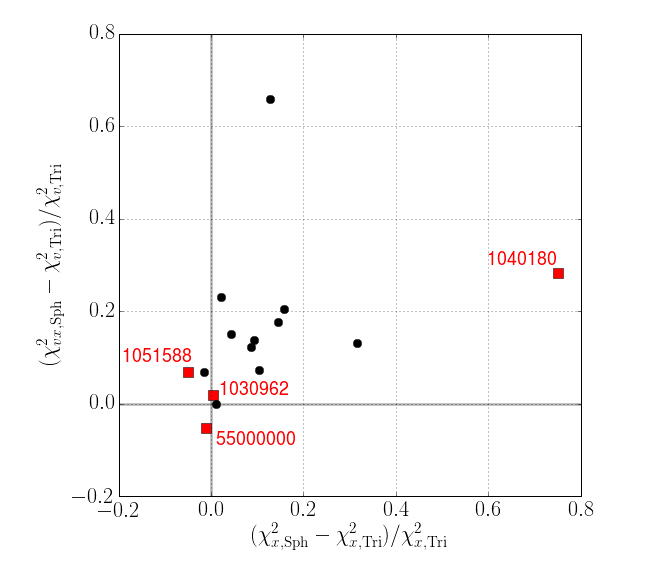}
\caption{Relative difference in ``chi-squared" (Equations \eqref{eq:chisq}) between the orbit and stream when the orbit is integrated in a spherical rather than triaxial potential. The x axis shows the difference in position, while the y axis shows the difference in velocity. The four highlighted streams shown as red squares are shown in several projections in Figure \ref{fig:streamsorbits1}; streams 1040180 and 1030962 are also shown in Figures \ref{fig:varymass1} and \ref{fig:varyr1}. }
\label{fig:chisq}
\end{figure}

\begin{figure*}[tp]
   \centering
   \includegraphics[width=0.85\hsize]{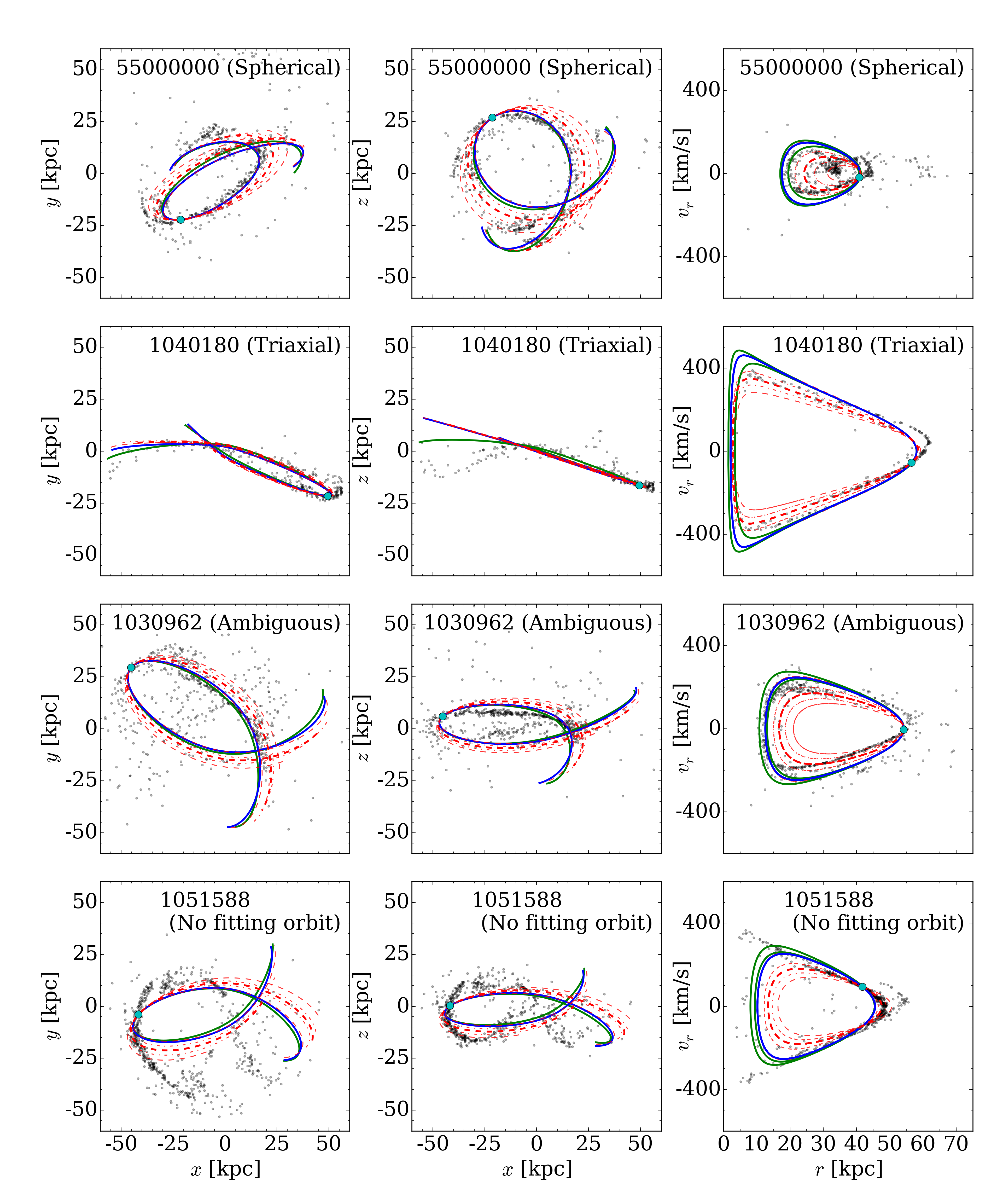}
   \caption{Results of integrating the orbit of the central particle (cyan filled circle) in the triaxial NFW potential (green solid line) and the spherical NFW potential (blue solid line), for 4 streams highlighted in Figure \ref{fig:chisq}. Both orbits were integrated with the ``true'' potential parameters derived as in S08. We also show the orbit integrated in the spherical NFW potential using our best-fit parameters, determined by maximizing the KLD (red long-dashed line), including error-bounds for the fit in $M_s$ (red short-dashed lines) and in $r_s$ (red dot-dashed lines). The stream stars are plotted in black. The row labels (``spherical,'' ``triaxial,'' etc.) indicate the potential shape with the lowest chi-squared for the stream pictured in that row. }
   \label{fig:streamsorbits1}%
\end{figure*}

\begin{figure*}[tp]
   \centering
   \includegraphics[width=\hsize]{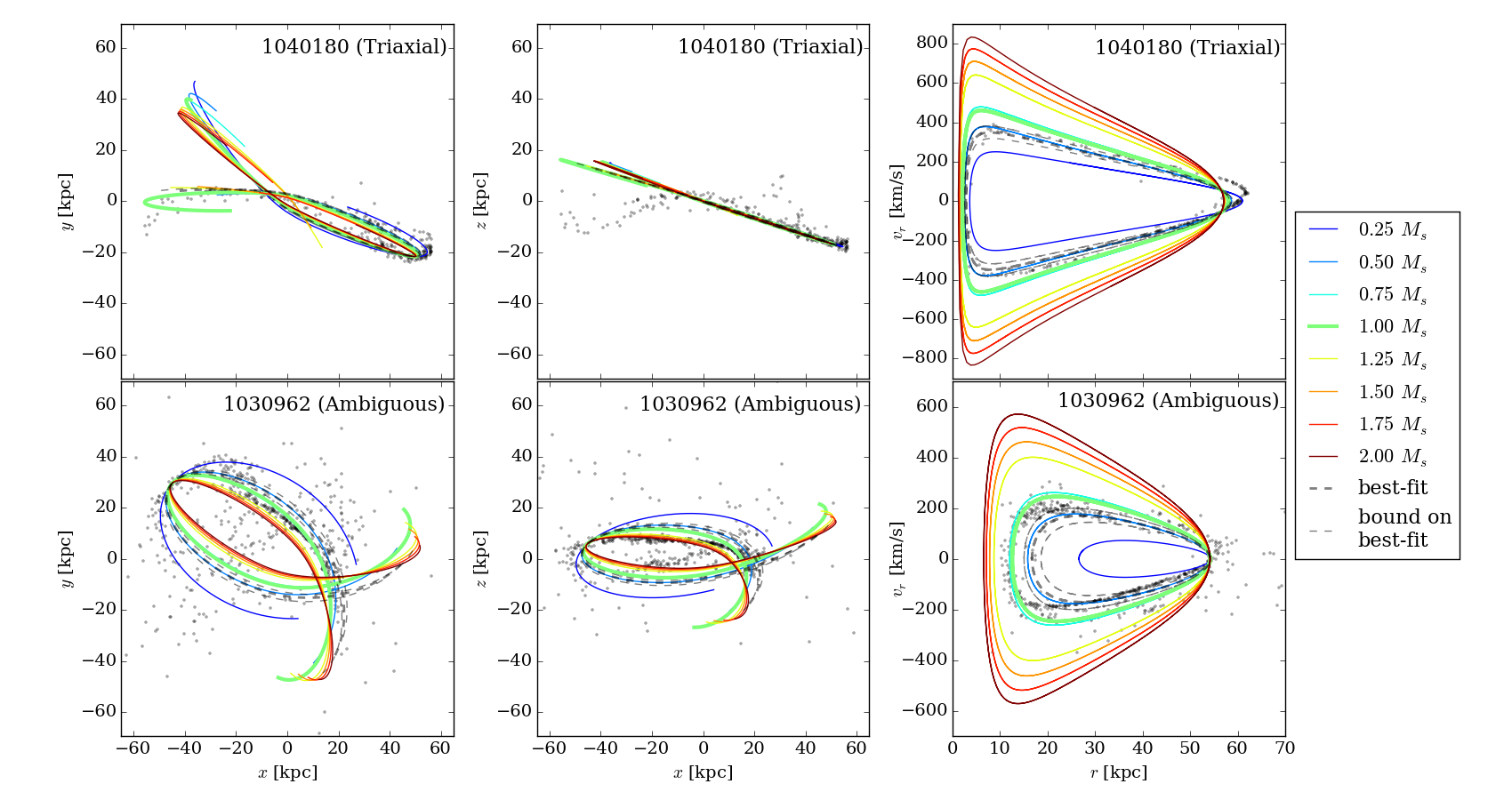}
   \caption{Integrated orbits of the central test particle in the spherical NFW potential for different values of $M_s$, fixing $r_{s,S08}=15.19$ kpc (see Table \ref{tab:par}), for an orbit where the triaxial potential is clearly better (top) and another that is more ambiguous (bottom). The orbit for $M_{s,S08}=1.87\times 10^{11} M_\odot$ from S08 is shown in green; the best-fit orbit according to the KLD is plotted in gray with error bounds in lighter gray. Other colors range from $0.25M_{s,S08}$ (dark blue) to $2M_{s,S08}$ (dark red) in steps of $0.25M_{s,S08}$.}
   \label{fig:varymass1}%
\end{figure*}

\begin{figure*}[tp]
   \centering
   \includegraphics[width=\hsize]{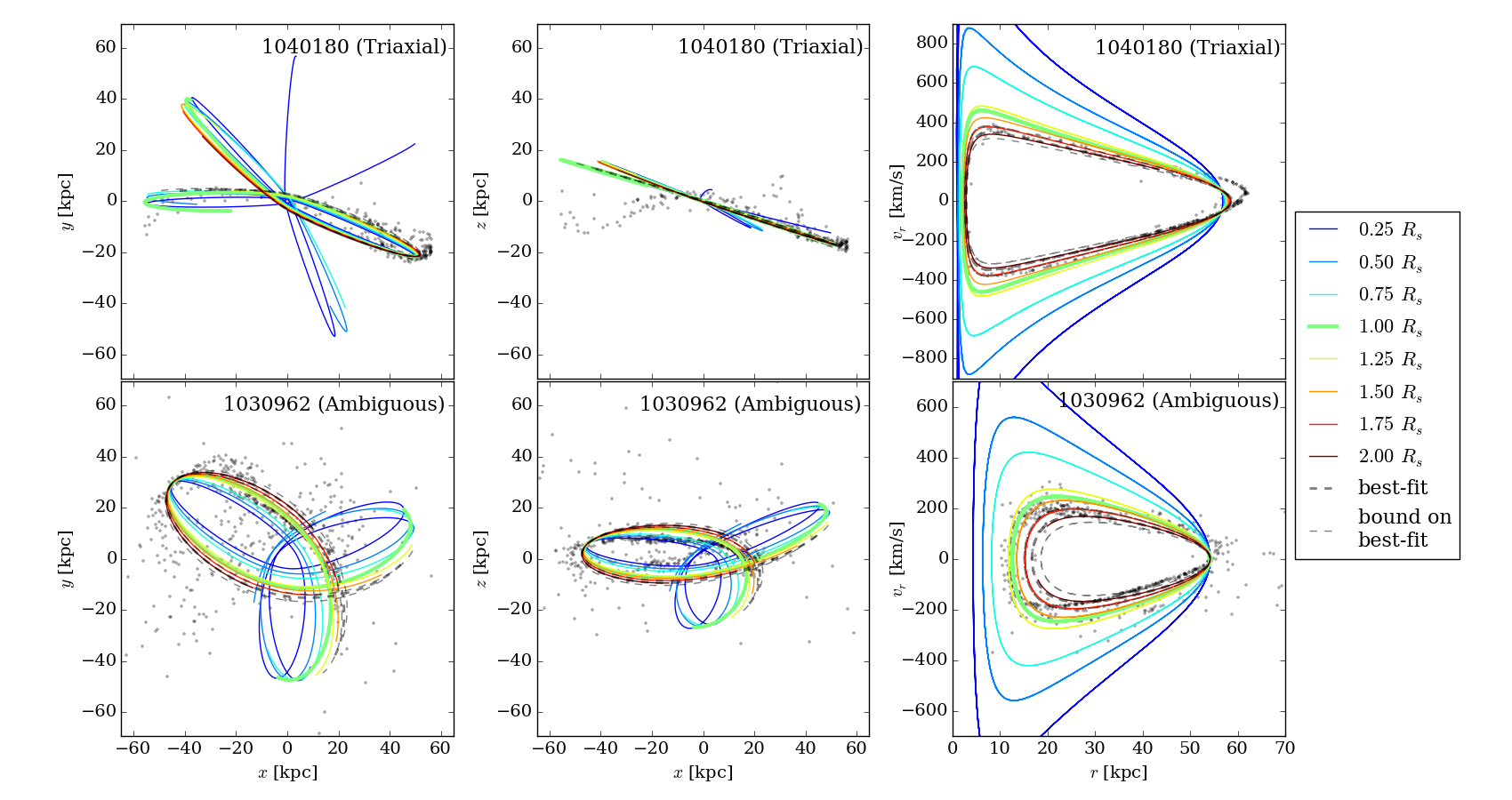}
   \caption{Integrated orbits of the central test particle in the spherical NFW potential for different values of $r_s$,  fixing $M_{s,S08}=1.87\times 10^{11}\ M_\odot$, for the same two streams as in Figure \ref{fig:varymass1}. The plotting scheme is the same as in Figure \ref{fig:varymass1}, but the colors now indicate a range of $r_s$ from $0.25r_{s,S08}$ (dark blue) to $2r_{s,S08}$ (dark red) in steps of $0.25r_{s,S08}$.}
   \label{fig:varyr1}%
\end{figure*}

We compared the alignment between the stars in each selected stream and the integrated orbits. We do not expect the streams to align exactly with the integrated orbits, not only because the spread of energies in the stream stars produces a slight misalignment, but also because the true potential in which the streams have evolved is both lumpy and time-evolving, as opposed to our smooth and static models. However, the goal is to see if the assumption of spherical symmetry makes a significant difference in how closely the integrated orbits follow the streams.

The values of the average minimum distances $\chi^2_x$ and $\chi^2_v$ (Equations \eqref{eq:chisq}) for all the streams in our sample are compared for the spherical and triaxial potentials in Figure \ref{fig:chisq}. The streams highlighted as red squares and labeled with their ID numbers in Figure \ref{fig:chisq} are shown in a few different projections in Figure \ref{fig:streamsorbits1}. In all projections, the stream stars are shown as black points; the central particle whose orbit is integrated forward and backward is marked with a big turquoise circle, and the integrated spherical and triaxial orbits are shown as blue and green solid lines, respectively. We also show for comparison the orbit integrated in the best-fit spherical potential determined with the KLD fit as a red dashed line, with accompanying red dot-dashed lines spanning the range of uncertainty of the fit.

 Figure \ref{fig:chisq} shows that in general there is a slight preference for a triaxial orbit over a spherical one in terms of average minimum distance between stars and orbit. There are two streams for which using a spherical instead of triaxial halo nearly doubles the average minimum distance; both these streams are on very radial orbits and the spherical orbit fails to incorporate precession out of the plane that is especially apparent in the  stars at the edges of the stream. One of these, stream 1040180, is shown in the second row of Figure \ref{fig:streamsorbits1}; the other stream has similar characteristics. Although (as expected) even the triaxial orbit does not perfectly line up with the stream edges, especially in the $x$-$z$ projection, it does a somewhat better job than the spherical orbit.
 
Other than these two outliers, for many of the streams it is hard to see by eye whether the triaxial or spherical halo is a better fit. Stream 1030962, shown in the third row of Figure \ref{fig:streamsorbits1}, is a good example of this situation. The orbits in the two potentials are nearly identical and the stream is much wider than the distance between the two orbits, and so it cannot discriminate. This stream, like most in the sample, shows some signs of discontinuity that also complicate the choice between potentials.
 
 Finally, one stream appears to slightly prefer the spherical potential, which is surprising given that we know the Aquarius halo is triaxial. This stream, 55000000, is shown in the first row of Figure \ref{fig:streamsorbits1}. Stream 55000000 is another case similar to 1030962, where the stream is much wider than the difference in orbits and small differences end up producing a chi-squared that slightly favors the spherical potential rather than triaxial in this case. 
 
 We can estimate the sensitivity of the chi-squared test using stream 1051588 (fourth row of Figure \ref{fig:streamsorbits1}), for which neither orbit lines up very well with the stream at all, so that in this case the differing chi-squared values are just choosing between two equally bad options. For this stream the chi-squared in position space favors a spherical potential while the chi-squared in velocity space favors a triaxial potential, but the fractional differences are only about ten percent. This illustrates that differences of this order in the chi-squared are not really indicative of a preference for one potential over the other.  On the other hand, differences on the order of 30 percent and larger in chi-squared, such as the differences shown by stream 1040180, are produced primarily when comparing the positions of the ends of the streams and orbits. This is consistent with the expectation that longer streams do a better job constraining the shape of the halo, since precession induced by departures from spherical symmetry has a more noticeable effect on a longer stream.

The influence of the halo mass on the orbit can be probed by varying the parameter $M_s$, as shown in Figure \ref{fig:varymass1} for two example streams, one where the triaxial potential is clearly a better fit (1040180, top row) and one where the two orbits are nearly indistinguishable from one another (1030962, bottom row). Increasing the halo mass shifts the orbit's apo- and pericenter inwards. The $r-v_r$ curve becomes more elongated for decreasing mass resulting in a larger radial range and a smaller velocity range covered. Looking at the projections onto different axes, orbits with $M_s \substack{+25\%\\ -50\%}$ all follow most of the stars in the stream. For reference, we also show the error bounds of the orbit using the best-fit parameters; these are typically a smaller range than we can distinguish by eye. 

We also probe the influence of the scale radius on the orbits, shown in Figure \ref{fig:varyr1} for the same two streams as in Figure \ref{fig:varymass1}. A larger scale radius results in a shift of apo- and pericentre towards larger radii and a more elongated radial velocity curve, which implies a larger radial range covered but a smaller range in radial velocities. Thus there is a partial degeneracy, in terms of orbital characteristics, between increasing the scale radius and lowering the scale mass (and vice versa). The orbit is generally less sensitive to $r_s$ than to $M_s$: even when the scale radius is increased to 1.75 times its measured value, the orbits still follow most of the stream. Thus, although the degree to which an orbit in a trial potential lies along a stream can select a rough range of potentials around the correct one, this method of determining the potential is neither very accurate nor very precise. We will show in the following section that a method that represents streams more realistically, as collections of stars on neighboring orbits, gives a better result in terms of both accuracy and precision.

\section{Potential fitting using action-space clustering}
\label{sec:KLD}
To fit a spherical NFW potential to the selected Aquarius streams, we use the method described in \citet{2015ApJ...801...98S}, which maximizes the clustering of the selected stars in the space of their actions, $\vect{J}$, by varying the potential parameters $\vect{a} \equiv (M_s, r_s)$ used to calculate the actions from the stars' positions and velocities. The potential parameters giving rise to the most clustered distribution of actions are chosen as the best fit, $\vect{a}_0$. 

\subsection{Measuring clustering with the KLD}
The amount of action-space clustering is measured statistically by calculating the Kullback-Leibler divergence (KLD),
\begin{equation}
 \sub{\super{D}{I}}{KL} =\int f_{\vect{a}}\left(\vect{J}\right) \log \frac{f_{\vect{a}}\left(\vect{J}\right)}{\super{f_{\vect{a}}}{shuf}\left(\vect{J}\right)}\ d^3\vect{J},
\label{eqn:KLDI}
\end{equation}
between the distribution of stellar actions for a specific set of potential parameters, $f_{\vect{a}}(\vect{J})$, and the product of its marginal distributions, $\super{f_{\vect{a}}}{shuf}(\vect{J})$. The product of marginals is constructed by computing the actions for a particular $\vect{a}$, then shuffling the different components of each action relative to one another to break correlations between actions, so we call it the ``shuffled" distribution. We use a modified Breiman density estimator \citep{2011A&A...531A.114F} to infer $f_{\vect{a}}$ and $\super{f_{\vect{a}}}{shuf}$ from the set of stellar actions, then calculate the KLD using numerical integration over a regular grid of $\vect{J}$, which replaces the integral in Equation \eqref{eqn:KLDI} with a sum over grid squares. More details on the numerical methods are available in \citet{2015ApJ...801...98S}. The larger the KLD, the more clustered the action space; the best-fit parameters are those for which the KLD is maximized.

Using the KLD as a figure of merit when fitting the potential has two advantages. First, because we measure clustering statistically there is no need to assign stars to a particular stream. Second, once the best-fit is found we can then use the KLD to set error contours on the best-fit parameters, $\vect{a}_0$, by comparing the action distribution for the best-fit parameter values, $f_{\vect{a}_0}(\vect{J})$, to the distribution for other trial values of the parameters, $f_{\sub{\vect{a}}{trial}}(\vect{J})$:
\begin{equation}
\super{\sub{D}{KL}}{II} \equiv \int f_{\vect{a}_0}(\vect{J}) \log \frac{f_{\vect{a}_0}(\vect{J})}{f_{\sub{\vect{a}}{trial}}(\vect{J})} d\vect{J}.
\end{equation}
This KLD is related to the conditional probability of the potential parameters $\sub{\vect{a}}{trial}$ relative to $\vect{a}_0$, averaged over the stars in the sample:
\begin{equation}
\super{\sub{D}{KL}}{II} = \langle \log \frac{\mathcal{P}(\vect{a}_0|\vect{J})}{\mathcal{P}(\sub{\vect{a}}{trial}|\vect{J})} \rangle_{\vect{J}}.
\label{eq:KLD2interp}
\end{equation}
A full discussion of this interpretation is in \citet{2015ApJ...801...98S}. Qualitatively, this expression measures how well the KLD can distinguish between the action distribution produced by the best fit parameters and the distributions produced by other parameters. Interpreting this as an uncertainty requires assuming that the distribution produced using the best-fit parameters is correct and comparing other distributions to it; hence the appearance of a conditional probability in Equation \eqref{eq:KLD2interp}.  As an example, if for some $\sub{\vect{a}}{trial}$ we get $\super{\sub{D}{KL}}{II}=1$, it means that those parameters are $e$ times less likely than the best-fit $\vect{a}_0$ to have produced the distribution of actions associated with the best-fit parameters (we are using natural logs everywhere). In a Gaussian probability distribution, (68, 95, 99) percent of the probability is inside the region where $\log P > -1/2 (-2, -9/2)$, so in this work we show the $\super{\sub{D}{KL}}{II}=1/2 (2,9/2)$ contours as rough analogs to one-(two-, three-)sigma uncertainties in the Gaussian case. However, analogies with Gaussian uncertainties should not be carried too far, since what our uncertainties really measure is the ability of the information in the action distribution to distinguish between different potentials, rather than the probability that the stream stars are drawn from a generative model of the action-space distribution (based on some potential parameters). Because we assume no generative model for the action-space distribution, the quoted uncertainties cannot be interpreted in a chi-squared sense. The level where e.g. $\log P = -1/2$  is more properly construed as the set of potential parameters that produce a distribution of stellar actions for which the probability that they are drawn from the most clustered distribution is $e^{-1/2} = 0.61.$ This interpretation takes into account 1) the unknown number of clumps in action space and their unknown positions, 2) the limited resolution of the distribution thanks to the finite number of stars in the sample, and 3) the way in which the action-space distribution changes as the potential parameters are changed.

\subsection{Computing the actions}
\begin{figure}[t]
	\centering
	\includegraphics[width=.8\hsize]{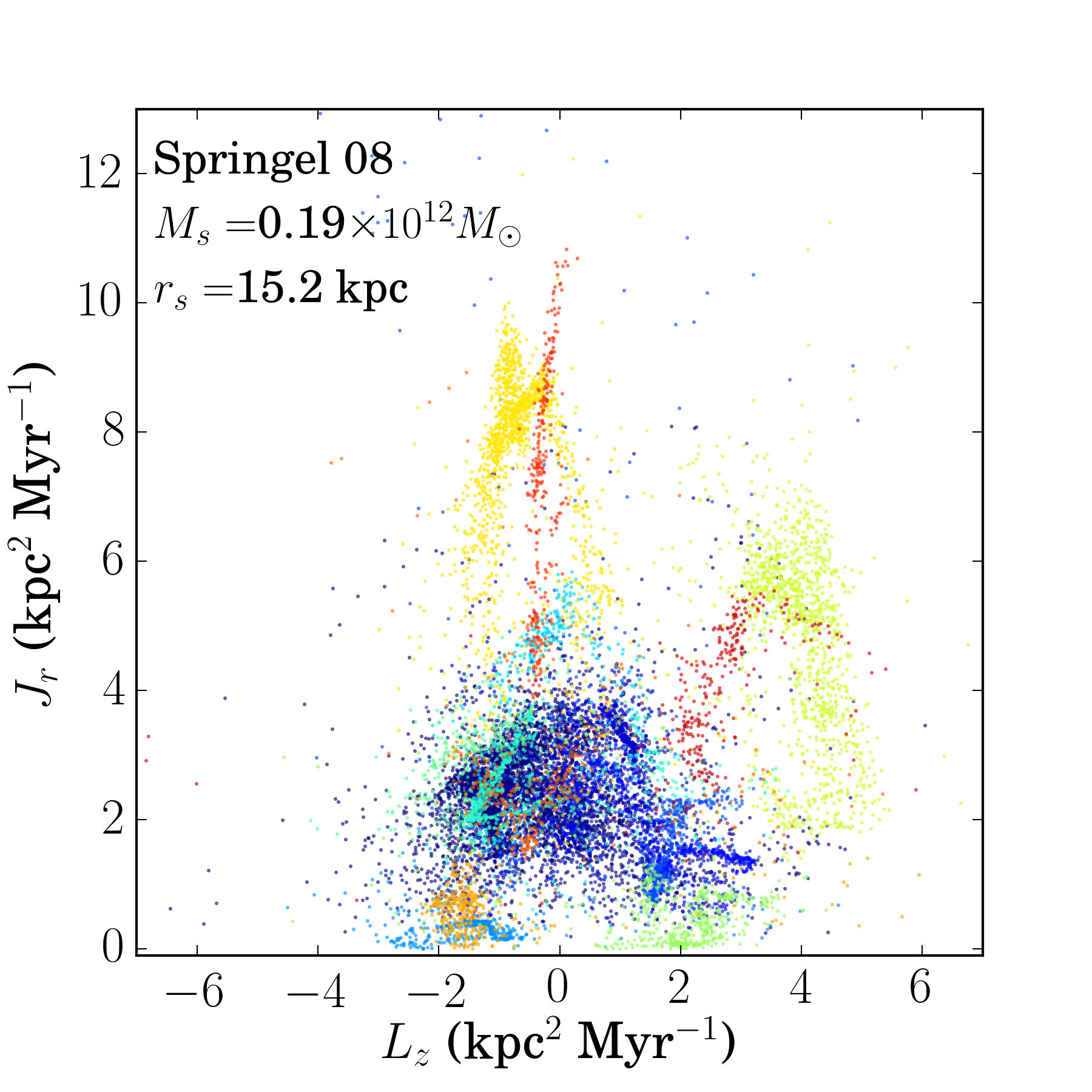}
	\caption{Distribution in action space of the streams in Figure \ref{fig:all_streams}, for the spherical NFW potential parameters derived from S08. The colors of the streams correspond to Figure \ref{fig:all_streams}.}
	\label{fig:actions1}
\end{figure}

A spherical potential has three independent actions that can be expressed in several different ways. We use the set comprised of the radial action $J_r$, the absolute value of the total angular momentum $L$, and the $z$ component of the angular momentum, which in our coordinate system points along the minor axis of the dark matter halo (though the potential to be fit is spherical). Thus our actions are $(J_r, L, L_z)$, of which only $J_r$ depends on the potential parameters. The other actions are included when calculating the KLD because although they do not change with the potential, they are still clumpy and correlated with $J_r$, so they improve the contrast between better and worse choices of the potential parameters. 

As when integrating the center-of-mass orbits, we use the potential of Equation \eqref{eq:PhiMsRs} to represent the spherical NFW halo, so our parameters $\vect{a}$ are the scale radius $r_s$ and the enclosed mass at the scale radius $M_s$. The angular momenta $L$ and $L_z$ are calculated from the stars' positions $\vect{x}$ and velocities $\vect{v}$:
\begin{equation}
\vect{L} \equiv \vect{x} \times \vect{v}; \qquad L \equiv |\vect{L}|; \qquad L_z \equiv \vect{L}\cdot\hat{z}.
\end{equation}
The radial action $J_r$ is calculated by numerical integration of
\begin{equation}
J_r = \frac{1}{\pi} \int_{\sub{r}{min}}^{\sub{r}{max}} dr\  \sqrt{2E - 2\Phi(r) - \frac{L^2}{r^2}} 
\end{equation}
where $r \equiv |\vect{x}|$ and the energy E is
\begin{equation}
E = \half \vect{v}\cdot \vect{v} + \Phi(r).
\end{equation}
The potential $\Phi(r)$, which depends on the parameters $r_s$ and $M_s$, is given by Equation \eqref{eq:PhiMsRs}. The integral endpoints $\sub{r}{min}$ and $\sub{r}{max}$ are determined by finding (also numerically) the two roots of
\begin{equation}
2E - 2\Phi(r) - \frac{L^2}{r^2} = 0.
\end{equation}
The range of $J_r$ varies as a function of the scale mass $M_s$, so to avoid undersampling and comparison issues when calculating the KLD (discussed further in Section 5 of \citeauthor{2015ApJ...801...98S}) we scale the radial action such that 
\begin{equation}
\label{eq:JrScaling}
\super{J_r}{scaled} \equiv \frac{J_r}{GM_s/\left(\ln 2 - 1/2\right)},
\end{equation}
which keeps the overall range of $J_r$ roughly constant, and comparable to the range of $L$ and $L_z$, for different $M_s$. 

Figure \ref{fig:actions1} shows the distribution of $(J_r, L_z)$ for the stars in the selected Aquarius streams, calculated using the values for $r_s$ and $M_s$ derived from $\sub{M}{200}$ and $\sub{r}{200}$ via the method outlined in S08 and Section \ref{ssec:potentials}. Although we are calculating the actions using a spherical approximation to the potential instead of the true triaxial one, we still see that both $J_r$ and $L_z$ are clumpy (as is $L$, not shown here) and that the clumps correspond to different streams (shown here in different colors, though the fitting method does not use this information). Thus the central assumption underlying our fitting method---that streams correspond to action-space clumps---is still satisfied.

\begin{figure}
	\centering
	\includegraphics[width=\hsize]{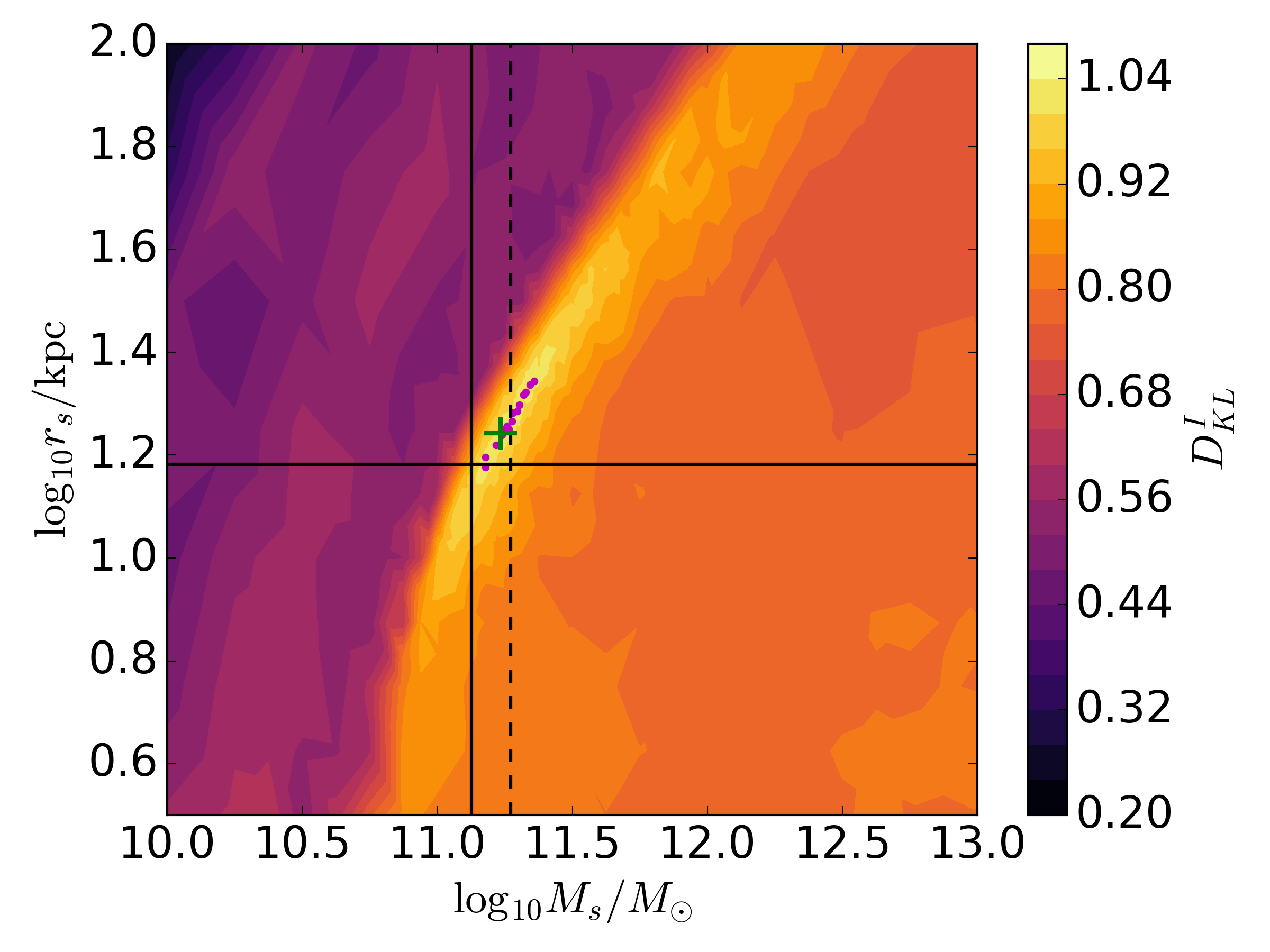}
	\caption{Contours of $\super{\sub{D}{KL}}{I}$.  The largest value of $\super{\sub{D}{KL}}{I}$ ($\vec{a}_0$) for the full sample is marked with a green cross; the small purple points show the best-fit values from leave-one-out samples (see Section \ref{ssec:uncert}). The dashed vertical line is the value of $M_s$ derived from S08, the solid vertical line is $M_s$ calculated from N10, and the solid horizontal line is the measured value of $r_s$ (the same in both papers; see Table \ref{tab:par}).}  
	\label{fig:kldMencl1}
\end{figure}

\subsection{Finding the best fit}

In order to find the best fit we compute $\super{\sub{D}{KL}}{I}$ (equation \ref{eqn:KLDI}) for a grid of parameter points, increasing the grid resolution adaptively in regions where the KLD is changing rapidly. We used 5 levels of adaptive refinement to converge the locations of the few highest KLD values. Figure \ref{fig:kldMencl1} shows the contours of $\super{\sub{D}{KL}}{I}$. We find that the best-fit parameters lie on a ridge of high $\super{\sub{D}{KL}}{I}$ with the very highest value (green cross) in between the scale mass derived from the N10 parameters and that derived from the S08 parameters. This is not surprising since the S08 values describe the mass profile best close to the virial radius (246 kpc), while the N10 parameters describe the mass profile better near the scale radius (15 kpc). Most of the ``stars" in our fitting sample are at distances somewhere in between these two radii, with an average distance around 40 kpc but reaching to about 120 kpc, so we expect our best-fit mass to interpolate between these two values. The scale radius value we obtain is slightly larger than the S08/N10 value; it is mainly determined by matching the enclosed mass at the average distance of the fitting sample, which gives rise to the degeneracy seen in the contours of Figure \ref{fig:kldMencl1}. 

\begin{figure}
	\centering
	\includegraphics[width=\hsize]{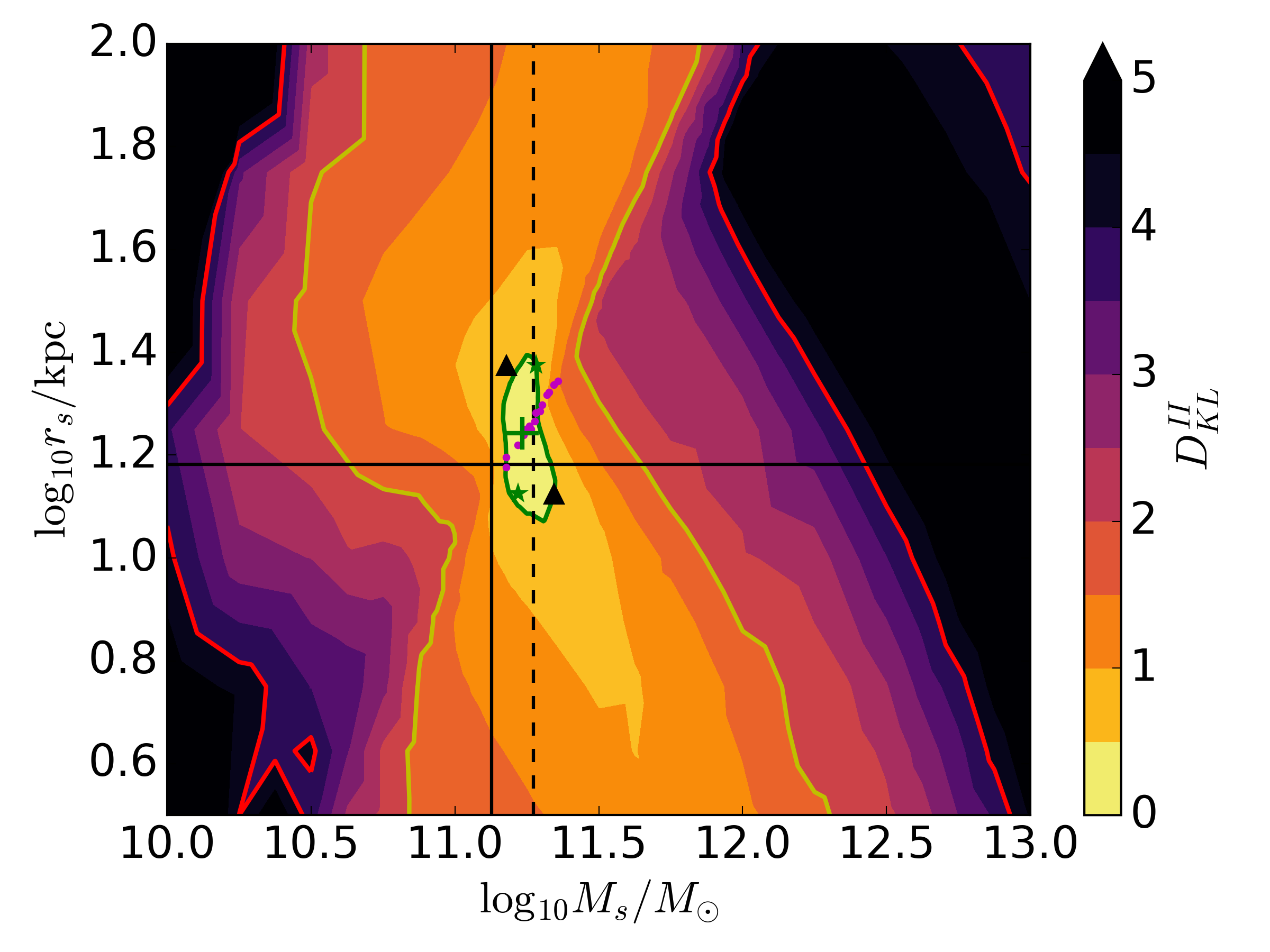}
	\caption{Contours of $\super{\sub{D}{KL}}{II}$. The red, yellow and green contours denote analogs to 68, 95 and 99\% confidence contours under the assumption of a roughly Gaussian probability distribution (as discussed in Section \ref{sec:KLD}) on the best fit value $\vec{a}_0$, shown as a green cross. The small purple points show the best-fit values from leave-one-out samples (see Section \ref{ssec:uncert}). The dashed vertical line is the value of $M_s$ derived from S08, the solid vertical line is $M_s$ calculated from N10, and the solid horizontal line is the measured value of $r_s$ (the same in both papers; see Table \ref{tab:par}). Black triangles indicate the extrema given by the one-dimensional uncertainties (red shaded region in Figure \ref{fig:mencl}); green stars indicate extrema taking into account the sense of the degeneracy between parameters in Step 1 (red dashed lines in Figure \ref{fig:mencl}).}
	\label{fig:kldMencl2}
\end{figure}

\subsection{Determination of uncertainties on the best-fit value}
\label{ssec:uncert}

\begin{figure}
	\centering
	\includegraphics[width = \hsize]{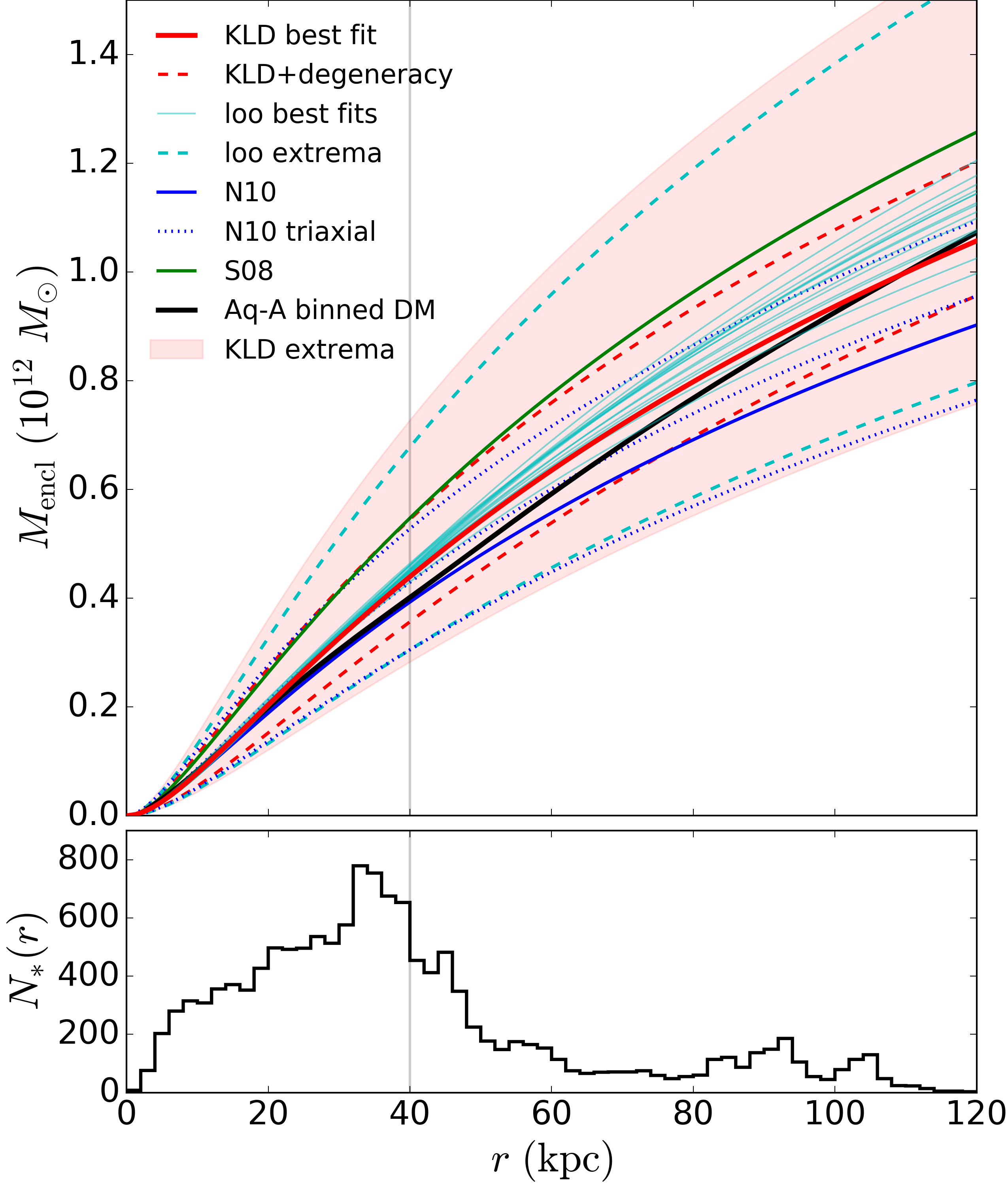}
	\caption{Best-fit enclosed mass profile (red solid line) and uncertainties predicted by the KLD (red) compared to the range of profiles obtained by leave-one-out analysis (cyan) as described in Section \ref{ssec:uncert}. The dashed cyan lines enclose the range of mass profiles obtained by taking the full range of $M_s$ and $r_s$ for all the leave-one-out fits. The red dot-dashed lines show profiles for the range of $r_s$ at best-fit $M_s$ based on the KLD, and the red dashed lines likewise show the range of $M_s$ at best-fit $r_s$. The shaded area encloses the full range of allowed profiles for the KLD-based uncertainties. The blue solid line shows the profile obtained by N10 (parameters in Table \ref{tab:par}) with the blue dotted lines the expected variation due to triaxiality as discussed in Section \ref{ssec:results}. The green solid line shows the profile obtained by S08 and the black solid line showed the spherically-binned mass profile calculated directly from the simulation. }
	\label{fig:mencl}
\end{figure}

  To determine uncertainties we use the KLD of Equation \eqref{eq:KLD2interp} to compare the distribution of actions at the points in the parameter grid with the one at the best-fit values identified in the first step in order to determine their error bounds, as described in Section \ref{sec:KLD}. In order to compensate for the different ranges in the radial action $J_r$ at different grid points in parameter space, we scale it with the prefactor of the potential, $G M_s (\ln 2 - 1/2)$, before comparing the action distributions with the KLD. Figure \ref{fig:kldMencl2} shows the contours of $\super{\sub{D}{KL}}{II}$ relative to the best fit (green cross). It is clear from these contours that the analogy with Gaussian uncertainties that motivated our choice of levels of $\super{\sub{D}{KL}}{II}$ is only plausible for $\super{\sub{D}{KL}}{II} \lesssim 1/2$ (green contour) and certainly not beyond that.

 To validate the use of Equation \eqref{eq:KLD2interp} and the choice of which value of  $\super{\sub{D}{KL}}{II}$ to use for setting uncertainties on the best-fit values, we performed a leave-one-out analysis on the set of 15 streams used for the fit. In practice this would not be possible, since we do not require membership information of stars in individual streams. For the same reason, this analysis should underestimate the uncertainty, since it does not fully account for our lack of assumptions about either stream membership or the number and locations of the streams in action space. However it does give a sense of the contribution that each stream makes to determining the best-fit model, which should drive the fit uncertainty.

We created 15 new samples by removing one of the 15 streams for each sample (presuming perfect membership knowledge), and re-ran step I on each sample to determine the best fit. We then compared the range of parameter values obtained in the leave-one-out fits to the range predicted for the full sample by Equation \eqref{eq:KLD2interp} for $\super{\sub{D}{KL}}{II}\leq1/2$. The different parameter values obtained by the leave-one-out fits are superposed as purple points on the full-sample contours of $\super{\sub{D}{KL}}{I}$ in Figure \ref{fig:kldMencl1} and $\super{\sub{D}{KL}}{II}$ in Figure \ref{fig:kldMencl2}.  Figure \ref{fig:mencl} compares the range of mass profiles obtained by the leave-one-out fits (cyan lines) with the range predicted by $\super{\sub{D}{KL}}{II}$ for the full sample (red lines and shaded region) and with several other ways of obtaining the mass profile. 

As seen in Figure \ref{fig:kldMencl1}, the leave-one-out fits select different points on a degenerate curve in parameter space, corresponding roughly to a constant enclosed mass at the mean radius of stars in the fitting sample, that is also followed by the contours of $\super{\sub{D}{KL}}{I}$ from the full sample. The primary difference between the range of parameter values predicted by Equation \eqref{eq:KLD2interp} and the range obtained from leave-one-out analysis, as shown in Figure \ref{fig:kldMencl2}, is that the KLD is insensitive to this enclosed-mass degeneracy when comparing neighboring action-space distributions. This is a reflection of the fact that in the first step of the analysis the KLD is used to estimate the degree of clustering at a given choice of parameters, which is sensitive to the fit degeneracy seen in Figure \ref{fig:kldMencl1}, while in calculating an uncertainty we use the KLD to compare action distributions at neighboring points in parameter space, which is insensitive to the degeneracy as is clear from the shape of the contours in Figure \ref{fig:kldMencl2}. It is also expected that the leave-one-out points will cluster more tightly than our fit results since we have implicitly used stream membership information to generate the different fitting samples. The range of $r_s$ values allowed by $\super{\sub{D}{KL}}{II}\leq1/2$ is also slightly larger than the range from leave-one-out, while the mass range is slightly narrower. This is an indication that the action-space distribution is more sensitive to changes in the mass parameter than the scale radius parameter, which is consistent with our results from orbit integrations in position and velocity space (Section \ref{ssec:orbits}). 

\begin{figure*}
\includegraphics[width=\textwidth]{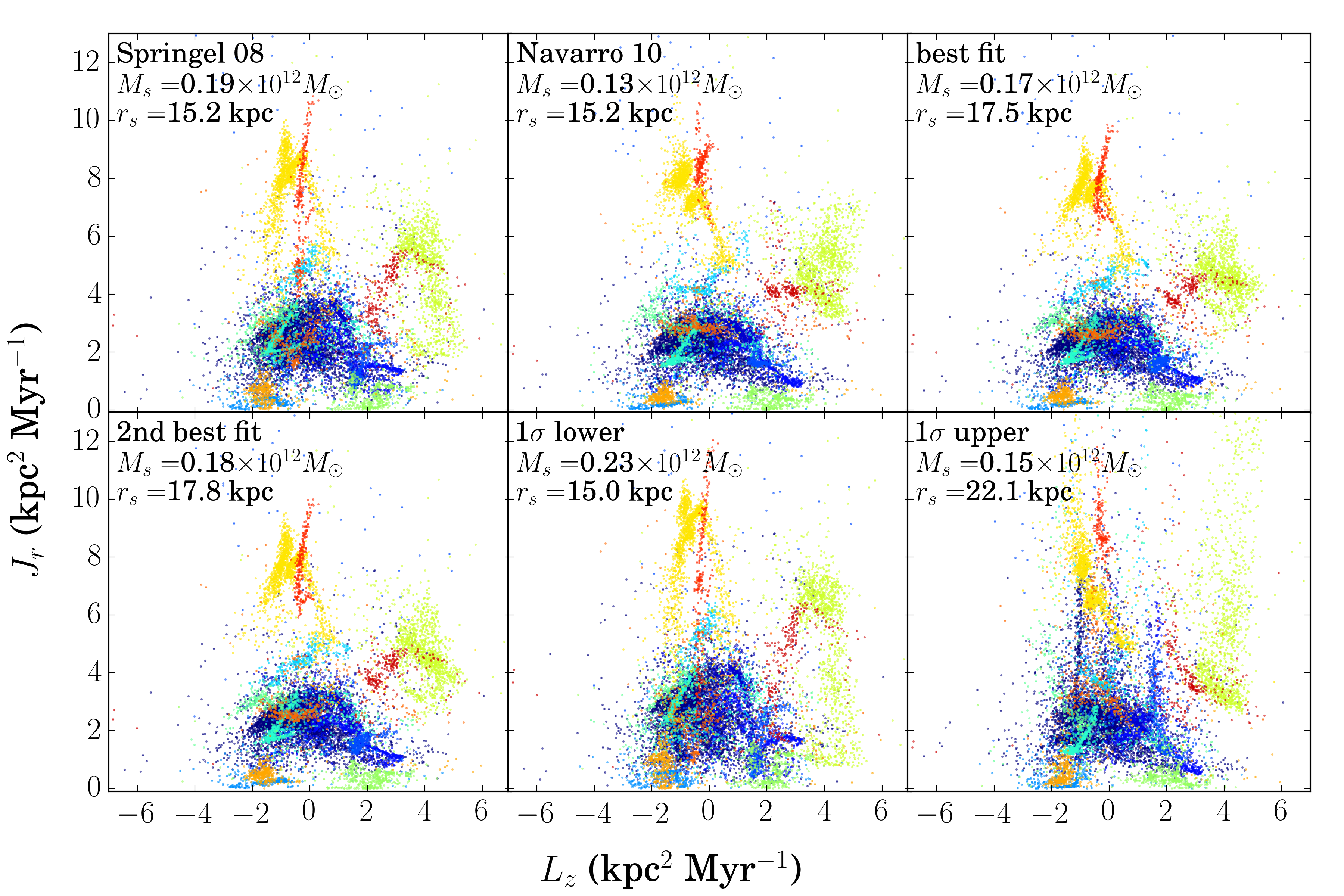}
\caption{Distribution of stream stars in action space for different parameter choices. From top left to bottom right: true potential parameters via S08, true potential parameters via N10, best-fit parameters, second best-fit parameters, and the two points marked with black triangles in Figure \ref{fig:kldMencl2}. In all panels the colors correspond to those in Figure \ref{fig:all_streams}.}
\label{fig:actions}
\end{figure*}

Figure \ref{fig:mencl} illustrates the range of allowed mass profiles resulting from different approaches to determining the uncertainty. The thick red line shows our best fit (the maximum value of $\super{\sub{D}{KL}}{I}$) for the full sample, while the thinner cyan lines show the results for the different leave-one-out samples. The red shaded region is the allowed range in the mass profile at the one-dimensional extrema of the green contour in Figure \ref{fig:kldMencl2}; that is, the upper limit is the mass profile with the maximum allowed value of $M_s$ and the minimum allowed value of $r_s$, while the lower limit is the profile with the minimum $M_s$ and maximum $r_s$. These points are marked as black triangles in Figure \ref{fig:kldMencl2}. Because this method of determining the uncertainty ignores the mass-radius degeneracy it produces a wider range of allowed profiles than the spread of the leave-one-out fits. If one takes the extrema of the best-fit parameters obtained by the leave-one-out fits in the same way, in the opposite sense from the mass-radius degeneracy (shown as thick dashed cyan lines) the range of allowed profiles is nearly as wide. Conversely, if we follow the sense of degeneracy outlined by the $\super{\sub{D}{KL}}{I}$ contours in Figure \ref{fig:kldMencl1} in choosing points on the $\super{\sub{D}{KL}}{II}\leq1/2$ contour (marked with green stars in Figure \ref{fig:kldMencl2}) then we get the range shown by the thick red dashed lines in Figure \ref{fig:mencl}, which is comparable to the spread in profiles from the leave-one-out analysis, although wider at small radii. The difference in the spread of allowable profiles reflects the difference in information between the leave-one-out analysis, which includes perfect membership assignments for all stars, and the KLD strategy, which does not use membership information at all. Our choice of $\super{\sub{D}{KL}}{II}\leq1/2$ as the uncertainty range is also supported by examining the difference between the best-fit distribution (top right panel of Figure \ref{fig:actions}) and the distribution generated by a point on the $\log P = -1/2$ contour (bottom right panel of Figure \ref{fig:actions}). The ``1-sigma'' distribution is visibly less clustered than the best-fit distribution.

This comparison indicates two possible routes to determining the range of allowed mass profiles without using membership information. The most conservative option, represented by the shaded red area in Figure \ref{fig:mencl}, is to report the full range of allowed parameter values in each dimension and allow all the combinations of parameters in that range as possible mass profiles. Slightly less conservative is to include information on the degeneracy between parameters obtained in the first step by reporting the range of allowed profiles following the sense of the degeneracy revealed by the contours of $\super{\sub{D}{KL}}{I}$. This approach will become more difficult as the mass model becomes more sophisticated.  

Combining the results of step I and step II the best-fit enclosed mass is $M_{s_0} = 1.72\substack{+0.57 \\ -0.20}\times 10^{11}\, M_{\odot}$ and the best-fit scale radius is $r_{s_0} = 17.46\substack{+4.60 \\ -2.48}$ kpc, where the one-dimensional uncertainties here and in Table \ref{tab:par} are the extrema of the 1-sigma contour marked as the two black triangles marked on Figure \ref{fig:kldMencl2}. The second best-fit mass is   $M_{s_1} = 1.81\times10^{11} \, M_{\odot}$ enclosed in $r_s = 17.78$ kpc. Since we find a larger scale radius than either S08 or N10, our value of the scale mass is closer to S08; however at their value of the scale radius (15.19 kpc) our best-fit mass profile has $M(R=15.19\textrm{ kpc}) = 1.43\substack{+0.90\\-0.52}\times 10^{11}\ M_{\odot}$, which is in between S08 and N10, and includes both values in its range of uncertainties.

\subsection{Results}
\label{ssec:results}

Figure \ref{fig:actions} shows the actions $L_z$ and $J_r$ at different points in parameter space. It can be seen that the KLD indeed recovers the most clustered action distribution: the actions are more clumped for the best-fit potential (upper right panel) than for the spherical potential with either set of parameters measured directly from the simulated halo (upper left and upper center panels). The lower two panels of the figure show that the actions get progressively less clustered when moving to parameter configurations farther from the best fit: the second-best fit (lower left panel) is visibly less clumpy at higher $J_r$, while a point on the $1\sigma$-equivalent contour produces a less clumpy distribution everywhere. 

Figure \ref{fig:mencl} shows the mass profiles for the different spherical halo models.  Our best fit from the KLD method is shown in red, with shaded error bounds showing the range of mass (dashed) and scale radius (dotted) within the red contour in Figure \ref{fig:kldMencl2}. The profiles using the parameters determined by S08 (green solid line) and N10 (blue solid line) bracket the empirical mass profile obtained by binning the dark matter particles in the smooth component of the simulated halo (black line). Our best fit is similar to the N10 fit at small radius where both agree with the spherically-averaged mass profile, but grows more steeply than N10 beyond about 20 kpc as does the spherically averaged profile. It never quite reaches the S08 mass profile, which agrees with the empirical mass profile close to the virial radius. We attribute this to the absence of stars in our fitting sample beyond 120 kpc (about half the virial radius of the halo).

We also show in Figure \ref{fig:mencl} the approximate mass profile along the three axes of symmetry in the triaxial halo (blue dotted lines with symbols), to give a sense of the degree to which the halo departs from spherical. More specifically, we take the average axis ratios of the \emph{density}, in the range 10-100 kpc, determined for Aq-A by \citet{2011Vera} in their Figure A2, obtaining $b/a = 0.7$ and $c/a = 0.55$. Renormalizing the axis lengths so that $a^2 + b^2 + c^2=3$ as in the Vogelsberger prescription for the triaxial NFW potential gives $a=1.29$, $b=0.91$, and $c=0.71$. We then plot three \emph{spherical} NFW mass profiles, using the mass and scale radius from N10, where the radius variable is rescaled by each of the three axis lengths: $\sub{M}{NFW}(r/a)$ (circles), $\sub{M}{NFW}(r/b)$ (triangles), and $\sub{M}{NFW}(r/c)$ (squares). These are not precisely the mass within isodensity contours along each symmetry axis, but do serve to give a rough sense of the variation of the mass profile between different directions in the halo. 

The S08 mass profile overestimates the mass compared to the empirical profile, which is (partly) due to the fact that subhalos are excluded from the empirical profile, but included when determining $r_{200}$ and $M_{200}$, which S08 use to set the profile parameters. N10 set their normalization at a much smaller radius, where most of the material is not in subhalos, and so are not as affected by the exclusion of bound substructure.

The best-fit mass profile traces the spherical NFW mass profile from N10 at small radii and the empirical mass profile at larger radii. We get the best agreement between our fit and the empirical mass profile at the radius at which we have the most stars, around $40$ kpc. The mass $M_s$ at the scale radius $r_s$ is marked with a diamond in the figure, corresponding to the crosshairs in Figures \ref{fig:kldMencl1} and \ref{fig:kldMencl2}. The KLD method recovers a larger scale radius $r_s$, and hence a larger enclosed mass $M_s$, than the profiles which were fit to the whole halo. The error bars on the fit are approximately as wide as the span between the profiles in S08 and N10, reflecting the inability of a single NFW profile to fit the Aq-A halo out to the virial radius. Compared to this difficulty, the effect of triaxiality is relatively small, as shown by the span of the mass profiles along the three principal axes in the triaxial potential.

\section{Discussion and Conclusion}
\label{sec:discussion}

In this work we investigated how assuming a smooth, time-independent potential with either spherical or triaxial symmetry affects the analysis of streams formed in a cosmological dark matter halo that is lumpy and time-evolving. First, we integrated the center-of-mass orbits for various streams in both spherical and triaxial potentials and compared how well the orbits traced the streams using the average minimum distance between the orbit and the stream stars (Section \ref{sec:integration}). We find that orbits integrated in a smooth, static potential resembling the present-day Aquarius A halo can trace the stellar streams extracted from the halo via stellar tagging, when starting from the position and velocity of a star midway through the stream. This agreement is striking given that the streams formed in a dynamic halo whose potential evolved with time and included many subhalos of various masses. In the majority of cases, the best agreement between orbit and stream was indeed achieved using a triaxial potential, which is not surprising since it is more representative of the true halo shape. However, in many cases the triaxiality of Aq-A was not enough to produce an appreciable difference (more than 10 percent in the average distance) between orbits integrated in spherical and triaxial potentials. For streams where we get good agreement with an orbit, the mass and scale of the halo can be roughly estimated by visually comparing the stream with a series of integrated orbits: the scale mass could be determined to within about 50\%, but the scale radius to within only about a factor of 2. Among the streams we compared, there are occasionally hints of the additional structure in the potential that was ignored; for example, in the one case where neither potential could produce an orbit that traced the stream, there appear to be gaps in the stream stars that might point to an interaction with a subhalo. However in general we find that the streams encode the present-day potential, and that ignoring substructure will not interfere catastrophically with the general tendency of streams to lie near orbits. This is similar to what \citet{2006ApJ...645..240P} found using numerical experiments with analytic potentials. Furthermore, as expected, the degree to which streams can distinguish between triaxial and spherical potentials via orbit fitting varies, depending on the progenitor's orbit and the extent of the stream. The most diagnostic streams in our sample were long and on very radial orbits; the extreme ends of these streams were most diagnostic in choosing a triaxial over a spherical potential.

Second, we tried fitting a smooth, spherical NFW mass profile to the entire set of 15 streams using the KLD method (Section \ref{sec:KLD}). We found that over the range of radii covered by stars in the fitting sample the best-fit profile followed the empirical, spherically-averaged profile computed directly from the dark matter distribution at the present day, roughly interpolating between the profile found by N10 that fits well at small radii and the profile found by S08 that fits well near the virial radius. The results confirm that this type of fit is insensitive to the adiabatic time-evolution of the host halo. This is expected, since the actions are adiabatic invariants; reassuringly, our fitting method has smaller uncertainties in $M_s$ and $r_s$ than simply comparing how well orbits line up with the streams by eye. Furthermore, stream-subhalo interactions in this model halo are not frequent or intense enough to destroy the action-space coherence of individual streams; neither is the poor assumption of a spherical rather than triaxial potential. In fact, our best-fit model produces a clumpier action distribution (and better agreement over a wider range of galactocentric distances) than two common ways of determining the same parameters directly from simulations: either by finding $M_{200}$ and $r_{200}$ (as in S08) or by determining $\rho_{-2}$ and $r_{-2}$ (as in N10). Although these two methods of parameterizing halos are initially independent of assumptions about the functional form of the potential, imposing the NFW functional form on either set of parameters to obtain $M_s$ and $r_s$ will only produce a good fit over a limited range of radii.  Our best-fit mass profile, which is effectively normalized around 40 kpc where we have the largest number of stream stars, agrees with the empirical profile from the simulation over a wider range of radii than either of the parameterizations, while also recovering $M_{200}$ within 10\% and $r_{200}$ within 4\% of the values determined directly from the dark matter distribution. Our difficulty fitting the Aq-A halo with an NFW profile is consistent with recent results by \citet{2015arXiv150700771H}, who find that the parameters they obtain using their fitting method are biased by up to 30\% for Aq-A because it differs significantly from the NFW form. 

We see indications that our model is not fully representative of the halo in the fact that there seem to be two sets of preferred parameters, according to $D_{KL}^{I}$ (Figure \ref{fig:kldMencl1}), that occupy a ridge in parameter space. The $M_s$ values of the two fits are nearly the same (and are within each others' approximate $1\sigma$ confidence contours) but the two preferred $r_s$ values differ substantially. Additionally, the uncertainty in $M_{200}$ for our best-fit model is comparable to the difference between normalizing the enclosed mass directly at $r_{200}$ and normalizing $\rho_{-2}$ at $r_{-2}$, and also to the variation of the mass profile along the different axes of symmetry thanks to its triaxiality, which was ignored in the fit. 

\citet[][hereafter B14]{2014ApJ...795...94B} explored the effect of assuming a smooth halo on the ability to determine the Milky Way's mass from fits to individual extremely narrow streams like those from globular clusters (GCs), evolved in the cosmological potential of the Via Lactea II simulation which, like Aquarius A, is both lumpy and time-evolving. They found that mass estimates from fits to single streams obtained using the streakline method \citep{2012MNRAS.420.2700K} could indeed be highly biased, and (as we do) that this bias was worse for streams closer to the Galactic center. They further found that assuming a smooth analytic potential limited the fundamental accuracy of mass estimates even when fitting many streams. Our work differs from this, and extends their results, in a few respects. 

First, the streams we study in this work more closely resemble those from small satellite galaxies than GCs: though their total stellar mass in some cases is similar, they tend to be much less concentrated in phase space initially than a GC would be. Second, the orbital distribution of streams within the Aquarius stellar halo is in our case determined by the cosmological simulation, whereas B14 inserted their GCs by hand at systematic locations. \citet{2010Cooper} show that the density profiles of the stellar halos derived from the Aquarius suite, and the luminosity-radius relation of the satellites that built up these halos, are consistent with observations of the stellar halo and satellites of the Milky Way, so we expect that the distribution and width of the Aquarius streams we study in this paper should reflect what is expected for the Milky Way. B14 find that mass estimates from streams beyond 70 kpc are significantly less biased, but the bulk of the stellar halo, and especially the part that will be observed by Gaia, is likely located at smaller galactocentric distances than this \citep{2000A&A...359..103R,2005ApJ...635..931B,Bell2008,2010Cooper}, so it is important to account properly for the radial distribution of stream stars when considering how well we will be able to determine the Galactic potential.
 
Second, we fit a limited sample of thin streams simultaneously rather than combining individual results, and use a method that does not require assigning stars to particular streams. We expect the action-clustering method to be more robust to scattering by lumpiness in the potential than methods, like the one used by B14, that compare distributions in 6D phase space. The effect of unaccounted-for lumps in the potential on the action-space distribution is mainly to dilute or subdivide the existing clumps of stars in a way uncorrelated between different streams.  This dilution will slightly lower the maximum information attained by the fit and thereby increase the size of the error contour, but because all streams are fitted simultaneously it should not introduce a bias in the final outcome. Motivated by this reasoning, as well as the number of narrow streams predicted by the Aquarius stellar halo model, we work with a much smaller number of streams than B14: 15 in our sample versus 256 in their case.

Our results demonstrate that fitting the potential using action-space clustering rather than comparisons in position- and velocity-space avoids some of the pitfalls illustrated by B14 in their paper. First, simultaneous fitting allows us to use relatively few cold streams and still recover $M_{200}$ within 10\%.  Second, analyzing the adiabatically invariant action-space distribution for goodness of fit, rather than making comparisons in position and velocity space, means we can use streams that are closer to the Galactic center: most of our stars are within 50 kpc and all are within 120 kpc, while B14 found that single streams closer than 70 kpc tended to have larger biases in the recovered mass. Finally, although we also find (like B14) that our assumptions lead to a slight under-estimate of the total mass, the true value lies well within our uncertainties, which reflect the degree to which our assumptions fail to match the true shape of the smooth potential rather than bias among individual streams. 

Most stream-fitting methods that have been tested to date, like the streakline method employed by B14 and its many recent variations \citep{2014MNRAS.443..423S,2014ApJ...795...95B,2015MNRAS.450..575A,2015MNRAS.452..301F}, require membership information for each stream to be fit and can therefore also preserve and use information about the angular phases of the stars in each stream, which is discarded by the KLD fit. Most of these tests have looked at the results of fitting individual streams and have found that single streams can produce either an excellent estimate of the potential or an extremely biased one, depending on the details of both the method used and the particular stream being fit. For example, \citet{2013MNRAS.436.2386L} found that fitting orbits to mock streams created in a smooth static halo produced estimates of the shape parameters and depth of a potential that were biased to about 20 percent, and that streams on different orbits placed different constraints on the potential parameters. Their results pointed toward using multiple streams fit simultaneously as we do here. Like our method, \citet{2013MNRAS.433.1826S} takes an action-space approach, but leverages the distinctive shape of the angle-frequency distribution of streams rather than their clumpiness in action space, by requiring that the slopes in angle and frequency be the same for a given stream in the correct potential. Their tests, using a single globular-cluster-like stream in a smooth static potential, show that without errors the circular velocity and shape parameter are recovered almost perfectly with this method. (Their tests used a scale-free isothermal potential, so were not subject to the mass-scale degeneracy we observe in our method.) However, they also find that the likelihood landscape is complex, with many local minima, and that introducing observational errors can create biases much larger than the formal uncertainties on the parameter estimates, though this can be improved by binning along the stream. They additionally find that the stream's length and overall orbital phase affects fit performance, with longer streams giving better constraints and streams at apocenter performing better than those near pericenter. These authors thus also argue for combining multiple streams to get a better fix on the potential. A more extended comparison of these different methods, and a treatment of the impact of using multiple streams, will be the subject of a forthcoming paper developed at the Gaia Challenge workshop series\footnote{\url{http://tinyurl.com/gaiachallenge}} (Sanderson et al. in prep).

Our results come with a few caveats. Although we do not need stream membership information to perform our fit, we did use it to select which streams to include in the sample, deliberately preferring thinner streams with a relatively narrow range of masses. This sort of sample is ideal for getting the best possible performance from our fitting algorithm since the action space has many tight clusters (and therefore high information) of similar size (so that smaller clusters are not overwhelmed). In this regard our sample is similar in nature to that used by B14, where the streams are all from a globular cluster model with the same mass and particle number. Furthermore we have not attempted to reproduce the proper number of stellar tracers or the expected observational errors, which will undoubtedly result in larger uncertainties and could also conceivably bias the fit results. 

However, our results do show that oversimplifying the model to be fit does not fundamentally produce a bias in the recovered mass profile; conversely, improving the model (for example, moving from a spherical to triaxial model) should reduce this contribution to the overall uncertainty. Our fitting method provides guidance on whether one model is a better representation than another: better models should be capable of producing a more clustered action distribution and hence a higher peak value of $D_{KL}^{I}$. We intend to test these two predictions in future work. 

\section{Acknowledgements}
RES is supported by an NSF Astronomy and Astrophysics Postdoctoral Fellowship under award AST- 1400989. AH acknowledges financial support from ERC Starting Grant GALACTICA-240271 and a NWO-Vici grant.



\bibliographystyle{apj} 
\bibliography{literature} 

\end{document}